*Gene Expression*

# A Fast and Flexible Method for the Segmentation of aCGH Data

Erez Ben-Yaacov[1], Yonina Eldar[1,*]

[1]Department of Electrical Engineering, Technion – Israel Institute of Technology, Haifa Israel.

**ABSTRACT**
**Motivation:** Array Comparative Genomic Hybridization (aCGH) is used to scan the entire genome for variations in DNA copy number. A central task in the analysis of aCGH data is the segmentation into groups of probes sharing the same DNA copy number. Some well known segmentation methods suffer from very long running times, preventing interactive data analysis.
**Results:** We suggest a new segmentation method based on wavelet decomposition and thresholding, which detects significant breakpoints in the data. Our algorithm is over 1,000 times faster than leading approaches, with similar performance. Another key advantage of the proposed method is its simplicity and flexibility. Due to its intuitive structure it can be easily generalized to incorporate several types of side information. Here we consider two extensions which include side information indicating the reliability of each measurement, and compensating for a changing variability in the measurement noise. The resulting algorithm outperforms existing methods, both in terms of speed and performance, when applied on real high density CGH data.
**Availability:** Implementation is available under software tab at: http://www.ee.technion.ac.il/Sites/People/YoninaEldar/
**Contact:** yonina@ee.technion.ac.il

## 1 INTRODUCTION

Array Comparative Genomic Hybridization (aCGH) is used to scan the entire genome for variations in DNA copy number. DNA from a test and reference cell populations is differentially labeled and hybridized on the array, and the log ratio between the two hybridization results is used to detect copy number variations. High density aCGH, spanning hundreds of thousands of probes, is a powerful tool in the research of cancer (Barrett *et al.*, 2004, Pinkel and Albertson 2005) and copy number polymorphisms (Conard *et al.*, 2006, Redon *et al.*, 2006, Perry *et al.*, 2008). A central task in the analysis of aCGH is the segmentation of the data into groups of probes that share the same DNA copy number.

Various segmentation methods have been proposed over the last years. Olsen *et al.* (2004) suggested a circular binary segmentation (CBS) algorithm, based on recursively applying a statistical test to detect significant breakpoints in the data. Picard *et al.* (2005) developed a dynamic programming procedure to segment the data when the number of segments is known in advance, which is referred to as CGHseg. The actual number of segments in real data is determined by maximizing a penalized likelihood function. While other segmentation methods exist, such as Lipson *et al.* (2005, 2006), a comparison study (Lai *et al.*, 2005) which tested 11 segmentation methods, concluded that CBS and CGHseg tend to have the best performance under various conditions. Another comparison study (Willenbrock *et al.*, 2005) compared 3 methods and proclaimed CBS as the method with the best results.

While presenting good segmentation performance, CBS is not sensitive to short segments, and often fails to detect them. On the other hand, CGHseg is sensitive to outliers in the data, leading to short segments corresponding to noise. A common drawback of both CBS and CGHseg is the long running time required to segment real high density arrays. Furthermore, it is not clear how to extend these methods to support side information.

In this paper we present *HaarSeg*, a new segmentation method, based on well known wavelet denoising principles. HaarSeg identifies statistically significant breakpoints in the data, using the maxima of the Haar wavelet transform, and segments accordingly. HaarSeg is a fast method, over 1,000 times faster than CBS and CGHseg, enabling interactive data analysis, with a slight compromise in performance. Due to its simple and intuitive structure, it is also a flexible method, and therefore easy to extend. We show how HaarSeg can be generalized to use quality of measurement data, additional information which exists in some platforms, indicating the reliability of each measurement. The use of quality of measurement was first suggested in Lipson *et al.* (2005), and it is currently used in ADM2, a segmentation algorithm based on Lipson *et al.* (2005, 2006) and used for example in de Smith *et al.* (2007) and in Perry *et al.* (2008). Since ADM2 does not have a freely available implementation we did not compare our performance to this segmentation algorithm. We also suggest an extension to compensate for the large variance in the log ratio measurements which occurs when one of the raw measurements has a very low value. Using these two generalizations, we show that HaarSeg outperforms existing methods, while remaining much faster.

The use of the Haar wavelet for microarray analysis is not new. Hsu *et al.* (2005) suggested applying standard wavelet denoising on microarrays, using the Haar wavelet. HaarSeg is different from that approach as it performs segmentation rather than smoothing of the data. To emphasize this difference we compare our results to Hsu *et al.*, and show that HaarSeg outperforms this method as well.

The rest of the paper is organized as follows: The basic HaarSeg algorithm is discussed in Section 2.1. Generalizations including quality of measurement data and adaptation to non-stationary variance are presented in Section 2.2 In Sections 3.1 and 3.2 we provide simulation results, and finally, analysis of real CGH data is presented in Section 3.3.

---

[*]To whom correspondence should be addressed.

## 2 METHODS

Each measurement in aCGH data is the log ratio of two raw measurements, red and green, which we denote by log(*R/G*).

Our signal, *y*[*n*], is a set of log(*R/G*) measurements from a single chromosome, ordered according to their genomic coordinates. Alternations in the number of copies in the aCGH data occur in contiguous regions of the chromosome, often spanning multiple probes. We therefore consider the problem of recovering a piecewise constant signal *x*[*n*] from its noisy measurements *y*[*n*], which can be viewed as the segmentation of *y*[*n*].

### 2.1 The Basic HaarSeg Algorithm

We suggest the following scheme, which is explained in detail in the next subsections:

- Apply the undecimated discrete wavelet transform (Mallat 1998) on the data, using the Haar wavelet.
- Select a set of detail subbands from the transform $\{L_{MIN}, L_{MIN+1},..., L_{MAX}\}$.
- Find the local maxima of the selected detail subbands.
- Threshold the maxima of each subband separately, using an FDR thresholding procedure.
- Unify selected maxima from all the subbands to create a list of significant breakpoints in the data.
- Reconstruct the segmentation result from the list of significant breakpoints.

*2.1.1 Undecimated Discrete Wavelet Transform*
The discrete wavelet transform (Mallat 1998) decomposes a given signal into an approximation subband and a set of detail subbands at different resolution scales. The approximation subband is a coarse or smooth version of the original signal, containing the scale coefficients. The detail subbands describe the higher frequencies of the signal, and are composed of the wavelet coefficients. Here we consider the undecimated discrete wavelet transform (UDWT), where each subband has the same number of coefficients. The UDWT is well suited for the task of data analysis, mainly due to its translation invariance property (Stark *et al.,* 2004).

The Haar wavelet is a natural choice for the recovery of piecewise constant signals (Mallat 1998). In this case, the detail coefficients of subband *L* are given by:

$$w_L[n] = \frac{1}{\sqrt{2^{L+1}}} \left( \sum_{k=n}^{n+(2^L-1)} y[k] - \sum_{k=n-2^L}^{n-1} y[k] \right). \quad (1)$$

The wavelet coefficients $w_L[n]$ in (1) can be viewed as the difference between two averages. In places where no breakpoint occurred in the signal, we expect $w_L[n]$ to be zero, as it is the difference between two identical averages. When zero mean additive noise is present it will typically average out for large enough *L*, so that $w_L[n]$ will still be close to 0. In places where a breakpoint occurred, we expect a high absolute value of $w_L[n]$, as the two averages are different.

Let $z_L[k]$ denote the local maxima of the absolute values of $w_L[n]$:

$$z_L[k] = \text{localmax}(|w_L[n]|), \quad 1 \leq k \leq K, \quad (2)$$

where *K* is the number of local maxima in $|w_L[n]|$. A coefficient is a local maximum if it is larger than its neighbors. We start by examining the two closest neighboring coefficients, and in case of equality we extend the neighborhood until we encounter a larger or smaller coefficient. High amplitude coefficients in $z_L[k]$ correspond to locations where abrupt changes occurred in *y*[*n*], and low amplitude coefficients correspond to changes in *y*[*n*] which were caused by noise. Finer detail subbands provide better localization of abrupt changes, but are more sensitive to noise.

*2.1.2 FDR Thresholding*
Given a list of coefficients *z*[*k*] from a specific subband *L*, we wish to keep just the larger ones, which in our case correspond to significant breakpoints in the data. To this end we consider the false discovery rate (FDR) thresholding procedure (Benjamini *et al.,* 1995), where FDR is defined as the proportion of false-positives out of all positives. FDR thresholding is a data-adaptive procedure, which controls the FDR. Specifically, we perform multiple hypotheses testing, where the null model assumes that the coefficient comes from a normal distribution with zero mean and a given standard deviation $\sigma$. We select the maximum number of coefficients such that the estimated FDR is kept under a predefined level *q*, where $0 < q < 0.5$.

To apply FDR thresholding we first sort *z*[*k*] in descending order, such that:

$$z_{(1)} \geq z_{(2)} \geq ... \geq z_{(i)} \geq ... \geq z_{(K)}.$$

For each measurement $z_{(i)}$ we calculate the two-sided p-value:

$$p_{(i)} = 2\left(1 - \Phi\left(z_{(i)} / \sigma\right)\right),$$

where $\Phi$ is the normal CDF. Starting from *i* = 1, we then find the largest index *i* for which

$$p_{(i)} \leq (i/K) q.$$

Thresholding is obtained by keeping the *i* largest coefficients, $z_{(1)}, ..., z_{(i)}$. Since in practice the standard deviation of the noise is unknown, we estimate it by using the robust median absolute deviation (MAD) estimator (Donoho 1995) on the finest detail subband $w_0[n]$;

$$\hat{\sigma} = \text{median}(|w_0[n]|)/0.6745. \quad (3)$$

*2.1.3 Signal Reconstruction*
To reconstruct the signal *x*[*n*] from the local maxima in each subband, we first need to unify maxima from all the selected detail subbands $\{L_{MIN}, L_{MIN+1},..., L_{MAX}\}$ into a single list of breakpoints. To take into account the possibility that the same breakpoint is detected at several levels with a slight offset, we use the following procedure. We first select all the significant coefficients detected at $L_{MIN}$, the finest detail level, and add them to the final list of breakpoints. We then add coefficients from level *L* = $L_{MIN}$ + 1, provided that they are at least $2^{L-1} + 1$ measurements away from any breakpoint in the final list. This step is repeated for all remaining subbands $L = L_{MIN} + 2,..., L_{MAX}$.

At the end of this process we remain with a single list of significant breakpoints in *y*[*n*]. Given the list of breakpoints, we estimate the piecewise constant signal *x*[*n*] by setting the value of the signal between two consecutive breakpoints to be the average of all probes in *y*[*n*] over that interval.

*2.1.4 Algorithm Parameters*
Two parameters need to be selected properly for HaarSeg:

*(1)* The set of detail subbands $\{L_{MIN}, L_{MIN+1},..., L_{MAX}\}$;
*(2)* The FDR parameter *q*.

The values of $L_{MIN}$ and $L_{MAX}$ are determined by the sampling resolution of our measurements. As $L_{MIN}$ increases, we are less sensitive to noise, but are also less likely to detect short segments in the data. As a general rule of thumb, if we expect a single segment in the data to span at least *k* probes, then we choose:

$$L_{MIN} = \lceil \log_2 k \rceil.$$

$L_{MAX}$ should be set large enough to reduce the sensitivity to noise, but small enough to avoid detection of slow, unimportant changes in the data, such as the genome-wide technical artifact described in Marioni *et al.* (2007). In all our experiments we used detail subbands {1, 2, 3, 4, 5}.

The FDR parameter $0 < q < 0.5$ controls the false discovery rate of breakpoints in the data. Low values of $q$ will reduce the false-positives at the possible cost of increasing the false-negatives, and vice versa.

### 2.1.5 Complexity

Let $N$ be the total number of measurements in $y[n]$. Calculating $w_L[n]$ in the case of Haar UDWT (1) can be performed in $O(N)$ operations regardless of the size of $L$, since it can be viewed as the difference between two running averages. FDR thresholding, applied to the transform maxima, has complexity $O(N\log N)$ as it requires sorting the data. Since the entire procedure is applied to a small finite set of detail subbands, the total complexity remains $O(N\log N)$.

## 2.2 Application to aCGH

We demonstrate the flexibility of HaarSeg by suggesting two extensions which are specific to aCGH. In Section 3 we show that these extensions lead to better segmentation on real aCGH data.

### 2.8.1 Quality of Measurement

Each raw measurement, red or green, is estimated from a set of pixels, associated with the same probe on the array. The median is usually used to estimate the raw measurement from the set of pixels. Current array platforms often provide the user with a value of $\sigma[n]$, which is the empirical standard deviation of the pixels corresponding to $y[n]$. High $\sigma[n]$ indicates poor measurement. The use of this additional information in a segmentation algorithm was first suggested in Lipson *et al.* (2005). This quality measure can be easily incorporated into our framework as well. Two steps need adjustment: the calculation of the wavelet coefficients and the final signal reconstruction.

The coefficients $w_L[n]$ in (1) can be rewritten as the difference between two averages:

$$w_L[n] = \sqrt{2^{L-1}} \left( \frac{1}{2^L} \sum_{k=n}^{n+(2^L-1)} y[k] - \frac{1}{2^L} \sum_{k=n-2^L}^{n-1} y[k] \right).$$

When each probe has a different variance, we suggest using the difference between two weighted averages for the calculation of $w_L[n]$:

$$w_L[n] = \sqrt{2^{L-1}} \left( \frac{\sum_{k=n}^{n+(2^L-1)} \frac{y[k]}{\sigma^2[k]}}{\sum_{k=n}^{n+(2^L-1)} \frac{1}{\sigma^2[k]}} - \frac{\sum_{k=n-2^L}^{n-1} \frac{y[k]}{\sigma^2[k]}}{\sum_{k=n-2^L}^{n-1} \frac{1}{\sigma^2[k]}} \right). \quad (4)$$

Note that when $\sigma[n]$ is constant for all $n$, (4) reduces to the original definition of $w_L[n]$ in (1).

To reconstruct the signal we use a weighted average instead, in order to estimate the signal values between two consecutive breakpoints:

$$\hat{\mu} = \sum_n \frac{y[n]}{\sigma^2[n]} \bigg/ \sum_n \frac{1}{\sigma^2[n]}.$$

### 2.8.2 Non-stationary Variance

In real CGH data, we observed that while most of the $\log(R/G)$ measurements have similar variance, there are segments of measurements with larger variance. Typically the raw measurements in those segments, either red or green, have a very low value compared to the rest of the raw measurements. An example from real data is shown in Figure 4. Note that in the previous subsection we discussed the variance of pixels inside the same probe, while now we consider the variance between consecutive probes. The connection between low value of the raw measurements and large variance of the log ratio can be explained by sensitivity analysis of the log ratio function:

$$\frac{\partial}{\partial R}\log(R/G) = \frac{1}{R}, \quad \frac{\partial}{\partial G}\log(R/G) = -\frac{1}{G}.$$

Hence, if all the raw measurements are perturbed with the same additive noise, then raw measurements with lower values will result in larger variations of the log ratio signal.

In the case of gene expression microarrays, several variance stabilization and normalization techniques have been suggested to cope with non-stationary variance. For example see the review of Steinhoff and Vingron (2006).

In order to adjust *HaarSeg* to reduce the effect of the non-stationary variance, we suggest splitting the transform peaks into two groups: a group of high variance, containing peaks that correspond to low raw measurements, and a group of typical variance that corresponds to the remaining measurements. We adjust the FDR thresholding to use these two variances accordingly, by suggesting the following scheme:

- Create a binary mask $b[n]$ using a fixed threshold $T_{NSV}$. Values of "1" correspond to probes with low raw measurements:

$$b[n] = \begin{cases} 1 & if \ \min(R[n], G[n]) < T_{NSV} \\ 0 & else. \end{cases}$$

- For each detail subband $w_L[n]$, defined in (1), calculate a matching binary mask $b_L[n]$. True values in $b_L[n]$ indicate that at least half of the measurements used to calculate $w_L[n]$ where marked as high variance in $b[n]$:

$$b_L[n] = \begin{cases} 1 & if \ \left( \frac{1}{2^{L+1}} \sum_{k=n-2^L}^{n+2^L-1} b[k] \right) \geq 0.5 \\ 0 & else. \end{cases}$$

- We estimate two standard deviations from the finest detail subband, $w_0[n]$, by splitting it to two groups according to the mask $b_0[n]$ and using the estimator in (3) on each group:

$$\hat{\sigma}_{high} \Leftrightarrow b_0[n] = 1$$
$$\hat{\sigma}_{typical} \Leftrightarrow b_0[n] = 0.$$

- Update the transform peaks $z_L[k]$, defined in (2), such that all the peaks will have the same standard deviation.

$$z'_L[k] = \begin{cases} z_L[k] / \hat{\sigma}_{high} & if \ b_L[k] = 1 \\ z_L[k] / \hat{\sigma}_{typical} & else. \end{cases}$$

- Apply FDR thresholding on $z'_L[k]$, using standard deviation of 1.

We set $T_{NSV}$ to a fixed value of 50 in our CGH analysis below.

## 2.3 Determining Aberrant Intervals

In the segmentation process of CGH arrays there is a need to determine which segments are aberrant, and set remaining segments to zero. As in CBS, CGHseg, and other segmentation methods, we approach this as a post-processing step. Several algorithms have been proposed for this task. A simple suggestion is to consider all segments with values outside *m* times the standard deviation range to be aberrant (Hodgson *et al.*, 2001), where *m* is frequently set to 3. An iterative method based on non-parametric statistical tests called MergeLevels was suggested in Willenbrock *et al.* (2005). Tibshirani *et al.* (2008) used an FDR based approach.

In our tests we used the simple method of considering all segments with values outside *m* times the standard deviation range to be aberrant. To estimate the standard deviation, we calculate the difference between $y[n]$, the original signal, and $x[n]$, the segmentation result, and apply the robust MAD estimator:

$$\hat{\sigma} = median\left(\left|y[n]-x[n]\right|\right)/0.6745.$$

Any other preferred method can be used instead, as this is simply a post-processing step.

## 3 RESULTS

We compared the performance of HaarSeg to CBS (Olshen *et al.*, 2004), CGHseg (Picard *et al.*, 2005), and to the wavelet denoising scheme suggested by Hsu *et al.* (2005), which we denote as Wave.

### 3.1 Simulated Data

In their comparison study, Willenbrock *et al.* (2005) created simulated CGH data using empirical distributions of segment length and copy number, taken from CBS segmentation results on real data. The noise model used in this simulation is additive i.i.d Gaussian noise. The original simulation contained 500 arrays, where every array included 20 chromosomes of 100 probes each. In order to simulate chromosome sizes which are closer to current high density CGH arrays, we modified Willenbrock's simulation to produce 100 arrays, each containing a single chromosome of 10,000 probes. We used the exact same model and noise levels used to produce the original simulations.

Since this simulation does not contain quality of measurement, or the original raw red and green measurements, we use only the basic HaarSeg algorithm, without any of the suggested extensions.

In order to compare results between HaarSeg and other algorithms, we computed the true positive rate and false discovery rate for all possible aberration thresholds, and plotted the receiver operating characteristic (ROC) curve for each segmentation algorithm. We computed the true positive rate (TPR) as the number of probes inside aberrations whose fitted values are above the threshold level divided by the number of probes inside aberrations. The false discovery rate (FDR) was calculated as the number of probes outside aberrations whose fitted values are above the threshold level divided by all the probes whose fitted values are above the threshold level.

The ROC curves and running times for HaarSeg, CBS, CGHseg and Wave appear in Figure 1. HaarSeg takes only 2 seconds to produce a result for all 100 arrays; this is over 1,500 times faster than CBS, and over 9,000 times faster than CGHseg, which was the slowest method. However, the speed gain of the basic HaarSeg algorithm comes with some performance price. HaarSeg performs slightly worse compared to CBS, about 1% worse in FDR and 1% worst in TPR. HaarSeg allows higher TPR than CGHseg, but at the cost of 1% in the FDR. Wave showed the worst ROC curve among the compared methods.

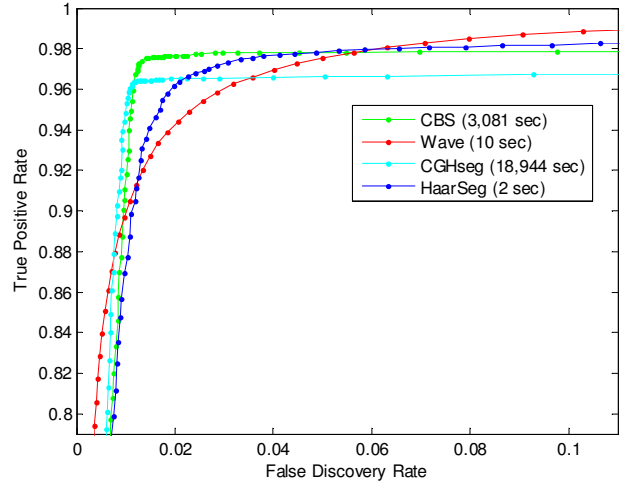

**Fig. 1.** ROC curves of the tested algorithms, using the simulation model from Willenbrock *et al.* (2005).

### 3.2 Simulated Data with Quality of Measurement

In order to test the performance gain when using our suggested extensions to HaarSeg, we created a simulation based on real data. We took 3 control self-self hybridization arrays, 236,404 probes each, from de Smith *et al.* (2007). These arrays contain quality of measurement and the raw red and green measurements. The true segmentation result of a self-self array is zero everywhere. We used the self-self arrays to create a simulation in the following manner: We reordered the self-self arrays and created 70 arrays of 10,000 probes each. For each array we created a mask of aberrant segments. Each segment was given a slightly different height, uniformly distributed between 0.1 and 0.2. To create the aberrant mask we used the empirical length distribution of CBS, taken from Willenbrock *et al.* (2005).

Figure 2 shows the ROC curve of TPR vs. FDR at various thresholds, and running times for all tested algorithms. We denote HaarSeg as the basic algorithm and W-HaarSeg as the algorithm with quality of measurement and non-stationary variance extensions described in Section 2.2. W-HaarSeg and CBS achieve the best results, where W-HaarSeg is about 1,000 times faster than CBS.

Using the empirical length distributions of CBS is biased towards CBS. Short segments of 2-4 probes rarely exist since CBS is not sensitive enough to detect such segments. We therefore repeated the experiment using the segment length distribution of W-HaarSeg, taken from segmentation results on the real data in de Smith *et al.* (2007). Since short segments are harder to detect, we increased the segment height to be uniformly distributed between the values of 0.15 and 0.25. Figure 3 shows the ROC curve and running times for this experiment. In this case, W-HaarSeg outperforms all the other tested methods. This demonstrates that W-HaarSeg is able to detect short segments, which CBS cannot, while keeping the false positive at a low rate.

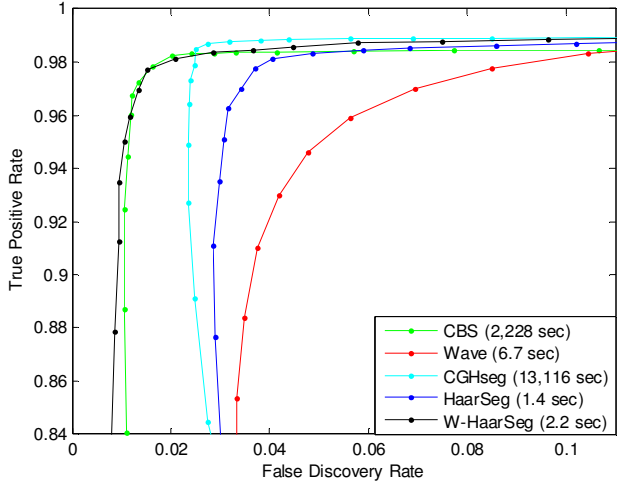

**Fig. 2.** ROC curves of the tested algorithms, using a simulation based on real self-self data, with segment length distributions of CBS.

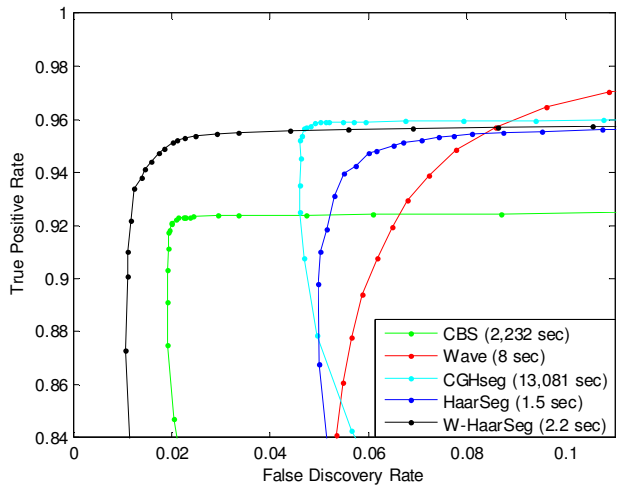

**Fig. 3.** ROC curves of the tested algorithms, using a simulation based on real self-self data, with segment length distributions of W-HaarSeg.

### 3.3 Real High Density CGH Data

In order to test performance on real high density arrays, we used data from de Smith *et al.* (2007), which enables performance evaluation to some extent, and contains quality of measurement side information. de Smith *et al.* (2007) compared samples from 50 healthy subjects to a reference sample in order to detect copy number variations in healthy individuals. This experiment also includes 3 control self-self hybridizations of the reference sample, used to estimate false positives.

Since the reference sample was a female and the test samples were males, we excluded chromosome Y from all our tests, and compensated chromosome X by adding a constant, estimated as the mean of the median of all X chromosomes in the 50 arrays. No other normalization was applied to the data. Each array therefore contains 23 chromosomes and a total of 236,404 probes. Each chromosome contains between 2,000-18,000 probes.

To estimate the FDR, we divided the average number of aberrant probes in the 3 self-self arrays, which we expect to be zero in the ideal case, by the average number of aberrant probes in the 50 arrays. Estimating the false negative is not possible on real data, where the exact true answer is not known.

We tested the performance of both HaarSeg, and W-HaarSeg, which is the HaarSeg algorithm with quality of measurement and non-stationary variance extensions described in Section 2.8. We compared results to CBS, CGHseg and Wave. For all tested segmentation methods, we used the aberrant threshold from Section 2.9, setting $m$ to 3.

Table 1 shows the FDR estimate, average number of active probes in the 50 arrays, and the time it took to segment all 53 arrays in each method. W-HaarSeg has the best false positive score, less than 1%, and CBS has the next best score, 4.3%. Compared to CBS, W-HaarSeg detects more active probes on average. This suggests that W-HaarSeg has a better false negative score, since it detects more probes, with a lower false positive estimate. Both HaarSeg and W-HaarSeg excel at running times compared to CBS and CGHseg. HaarSeg and W-HaarSeg segment the entire data in less than one minute, while CBS takes 10 hours and CGHseg takes 66 hours to produce the segmentation result.

**Table 1.** Results for real data

| Method | FDR | Avg. active probes num. | Run time |
|---|---|---|---|
| CBS | 4.3 % | 4603 | 36,420 sec |
| CGHseg | 10.0 % | 5031 | 237,600 sec |
| Wave | 10.7 % | 6284 | 121 sec |
| HaarSeg | 9.9 % | 5317 | 29 sec |
| W-HaarSeg | 0.9 % | 4782 | 38 sec |

Figure 4 demonstrates the non-stationary variance effect in a section from a self-self array. The correct segmentation result in this case is zero everywhere. Only W-HaarSeg achieves an exact zero result for this section.

Figure 5 shows an example of segmentation results of a short possible deletion spanning 4 probes. The true answer is not known, but in this example CBS was the only method that did not detect the deletion, indicating that CBS is less sensitive in the detection of short segments. This example also demonstrates the difference between the results of Wave, where each measurement has a different value, and HaarSeg, where all measurements in the same segment share the same value.

### 3.4 Parameter Settings

We used R package DNAcopy version 1.12 for CBS, R package tilingArray version 1.16 (Huber *et al.*, 2006) for CGHseg, and R package waveslim version 1.6.1 (Whitcher 2007) for Wave.

We used default parameters for CBS. For CGHseg we set the maximum number of allowed segments in a chromosome to 300 and the maximum length of a segment to 2,000 probes. To determine the actual number of segments in CGHseg we used the BIC penalty term. For Wave we used SURE soft thresholding with a maximum detail subband of 4, according to the description in Hsu *et al.* (2005). For HaarSeg, we used 5 detail subbands, $L =$

{1,2,3,4,5} and set *q* to 0.05 for the simulated data in Section 3.1, and $q = 0.001$ for the simulated data in Section 3.2 and for the real data in Section 3.3. Running times were calculated on AMD Athlon 64X2 with 2GB RAM.

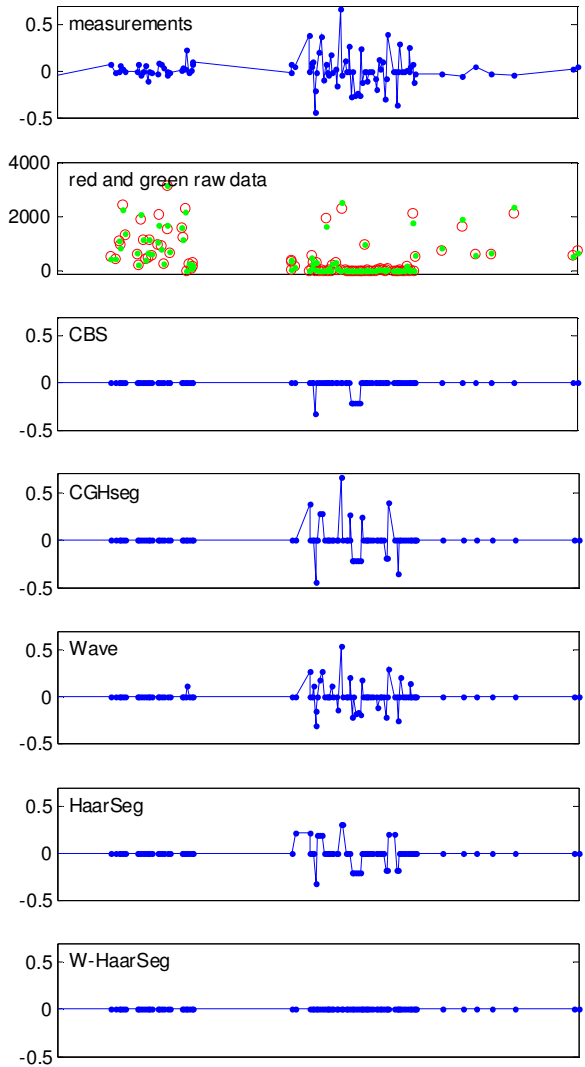

**Fig. 4.** Segmentation results of a section from chromosome 1 in a 'self-self' array, GSM215042, demonstrating the non-stationary variance effect. Graphs are in genomic coordinates. The correct result is zero at all the probes. Segmentation results are shown after applying the aberration threshold.

## 4 DISCUSSION

We presented HaarSeg, a new method for the segmentation of high density aCGH. Applied on both simulated and real data, our method is considerably faster, but with a slight performance penalty compared to leading approaches. We demonstrate the flexibility of our method by suggesting two extensions. First, we propose using quality of measurement. This additional information, when it exists, enables HaarSeg to better handle outlier measurements. Second, we suggest an extension to compensate for the large vari-ance in part of the log ratio measurements, which occurs when at least one of the raw measurements has a very low value. This extension enables HaarSeg to avoid over segmentation. Using both additions, HaarSeg outperforms existing algorithms.

It is interesting to note that each of the two suggested extensions contributes about the same performance gain to the final result. Applying just one of the extensions, either the quality of measurement or the non-stationary variance, will result in about half the total performance gain. These extensions do not change the low complexity of HaarSeg, and running times remain short. The importance of reasonable running times will become more and more evident as microarray size and resolution continue to grow rapidly.

While we showed application of our method to aCGH, where we seek to detect breakpoints in the data, our method can also be extended to detect other interesting features in microarray data. This is a subject for future research.

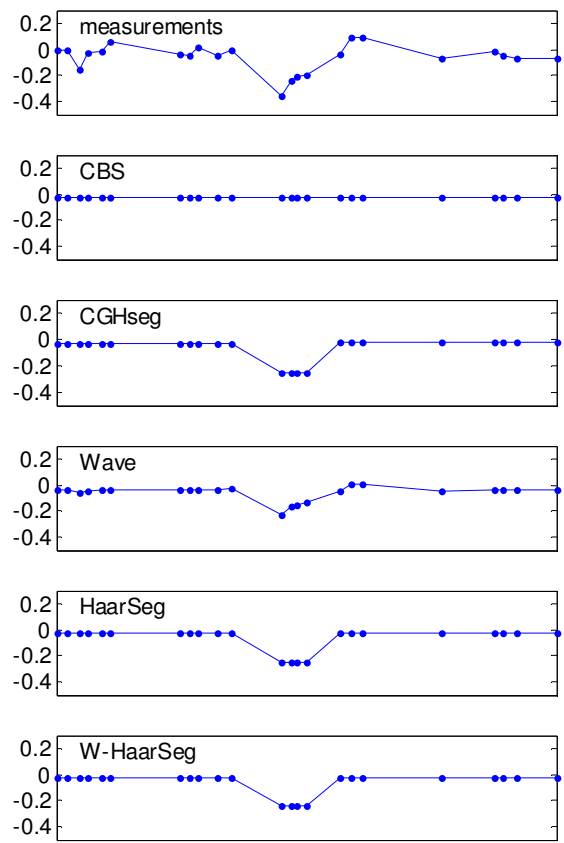

**Fig. 5.** Segmentation results of a possible deletion in chromosome 6, array GSM214509. Graphs are in genomic coordinates. Segmentation results are shown before applying the aberration threshold.

## ACKNOWLEDGEMENTS

The authors would like to thank Prof. Zohar Yakhini for suggesting the use of quality of measurement, for many fruitful discussions regarding aCGH data, and for useful comments on the paper. The authors would also like to thank Anya Tsalenko and Adam J. de Smith for their assistance with the real data presented in this